\documentclass[12pt,british,a4paper]{article}
\usepackage{cite}
\usepackage{babel}
\usepackage{graphicx}
\usepackage{amsmath,amsfonts}
\newcommand{\half}{\frac{1}{2}}
\newcommand{\Tr}{\mathrm{Tr}}

\def\ie{\textit{i.e.}}

\def\IR{\mathbb{R}}

\begin{document}
\begin{flushright}
DCPT--04/33
\end{flushright}
\bigskip
\begin{center} 
  {\Large \bf The Exact Geometry of a}
  \bigskip
  {\Large \bf Kerr-Taub-NUT Solution of String Theory }
\end{center}

\bigskip\bigskip\bigskip

\centerline{\bf Harald G. Svendsen} 

\bigskip\bigskip\bigskip
\centerline{\it Centre for Particle Theory} 
\centerline{\it Department of Mathematical Sciences} 
\centerline{\it University of Durham} 
\centerline{\it Durham, DH1 3LE, U.K.}
\centerline{\small \tt h.g.svendsen@durham.ac.uk}

\bigskip\bigskip
\abstract

In this paper we study a solution of heterotic string theory
corresponding to a rotating Kerr-Taub-NUT spacetime.
It has an exact CFT description as a
heterotic coset model, and a Lagrangian formulation as a gauged WZNW
model. 
It is a generalisation of a recently discussed stringy
Taub-NUT solution, and is interesting as another laboratory for
studying the fate of closed timelike curves and cosmological
singularities in string theory.
We extend the computation of the exact metric and dilaton to this
rotating case, and then discuss some properties of the metric, with
particular emphasis on the curvature singularities.

\newpage


\section{Introduction}


Recently, the exact geometry was computed
\cite{Johnson:2004zq} in a Taub-NUT \cite{Taub:1951ez,Newman:1963yy}
type solution of heterotic string theory, which led to interesting
observations regarding closed timelike curves and cosmological
singularities. 
The computation was done by realising the solution as an exact
conformal field theory (CFT) given as a heterotic coset model
\cite{Johnson:1995ga,Johnson:1995kv}, and using a Lagrangian
formulation in terms of a gauged Wess-Zumino-Novikov-Witten (WZNW)
model. In this description, it is relatively easy to write down the
effective action, which takes into account all $\alpha'$ corrections
and therefore allows us to extract the exact solutions for the
spacetime fields.  
The low-energy limit \cite{Johnson:1994jw} of this solution is known
to be a special case of a larger family of solutions which as well as
a NUT parameter $\lambda$ also has an angular momentum parameter
$\tau$ \cite{Galtsov:1994pd}, and is referred to as the stringy
Kerr-Taub-NUT spacetime. 
This more general solution was related to a heterotic coset model in
ref.~\cite{Johnson:1995ga}, where it was demonstrated that the
low-energy limit of the coset model corresponds to the throat +
horizon region of the stringy Kerr-Taub-NUT spacetime, in a way
analogous to the non-rotational case
\cite{Johnson:1994ek,Kallosh:1994ba}. 

Having thus another example of an exact CFT which can be described as
a gauged WZNW model, it is interesting to carry out the same analysis
as in ref.\cite{Johnson:2004zq}, and deduce the exact spacetime
fields. The aim of this paper is to do exactly this, and then to
discuss some properties of the exact metric.

The construction of the heterotic coset model, and the computation of
the spacetime fields are very analogous to what was done in
ref.\cite{Johnson:2004zq}, so we will only briefly summarise this in
the following. 
The study of this paper goes beyond that of ref.\cite{Johnson:2004zq}
in that we now give up spherical symmetry (for $\tau\neq 0$), and the
main purpose of this paper is to discuss some of the effects the
rotation has on the spacetime. 

\bigskip\noindent
There is a scarcity of known exact solutions of string theory which
has obstructed a better understanding of the theory beyond the
supergravity limit.
In this respect it is worthwhile to study any examples of exact
solutions that can be found.
The known solutions fall into three primary categories. First are
Minkowski space, and orbifolds of Minkowski space. Second are plane
wave solutions. These two classes of solutions receive no $\alpha'$
corrections due to special properties of the background.
Third are gauged WZNW models, which is the type of exact solutions to
be discussed in this paper. These solutions are exact by virtue of
having an exact worldsheet CFT description.

The stringy Kerr-Taub-NUT spacetime we shall investigate has closed
timelike curves (CTCs) as well as cosmological Big Bang and Big Crunch
singularities in the low-energy limit. This model therefore provides a
laboratory for studying the fate of CTCs and cosmological
singularities in string theory.

\bigskip\noindent
We find that the horizons connecting the Taub and the Kerr-NUT regions
of the spacetime are independent of the the high-energy
corrections. The same is true for the 
outer boundary of the ergosphere (the boundary of stationary motion). 
In the neighbourhood of these regions the
spacetime is modified, but in a mild way that does not alter the basic
features of the spacetime.
In particular, the CTCs and the cosmological singularities are there
just as in the low-energy approximation.
This result is completely analogous to the non-rotating case.

The curvature singularities, on the other hand, are significantly
shifted. As in the non-rotation case, a new Euclidean region appears,
whose nature, however, depends on the rotation.
If the rotation is small this region appears as a shell surrounding the
horizon in one of the NUT regions, while if the rotation is large,
there are Euclidean ``bubbles'' outside the horizons in both NUT
regions.
However, it should be kept in mind that near the curvature
singularities, string loop effects become important as the dilaton
blows up.

\section{Exact conformal field theory}

The model we shall study is an exact conformal field theory by virtue
of having a formulation as a coset model based on the coset
$G/H=SU(2)\!\times\!SL(2,\mathbb{R})/U(1)_A\!\times\!U(1)_B$.
This construction gives no reference to a background spacetime, which
is therefore considered as a derived concept. What makes coset models
particularly interesting is that they also have a Lagrangian
formulation in terms of gauged
Wess-Zumino-Novikov-Witten (WZNW) models.
The WZNW action is written in terms of a bosonic field
$g(z,\bar{z})\in G$ which is a map from the worldsheet $\Sigma$ to a
Lie group $G$. Explicitly,
\begin{equation}
  I_{WZNW}=-\frac{1}{4\pi}\int_\Sigma d^2\!z\Tr(g^{-1}\partial g
  g^{-1}\bar{\partial}g) + i\Gamma(g),
\end{equation}
where
\begin{equation}
  \Gamma(g)=\frac{1}{12\pi}\int_\mathcal{B}\Tr(g^{-1}dg)^3
\end{equation}
is the Wess-Zumino term, and $\mathcal{B}$ is a three-dimensional
space whose boundary is $\Sigma$.
A WZNW model is then governed by an action $S=kI_{WZNW}$, where the
constant $k$ is referred to as the level constant. By comparing this
action with the nonlinear sigma model, it is easy to see that we
should identify this constant with the string tension, \ie\ 
$k\sim \frac{1}{\alpha'}$.

The coset $G/H$ corresponds in the Lagrangian formulation to gauging
the subgroup $H$.
To make a gauged WZNW model, we introduce a gauge field
$A\in\text{Lie}(H)$, and replace the derivative with a covariant
derivate in the first term of 
the WZNW action. This gives a gauge-invariant term. The WZ
term on the other hand has no gauge-invariant extension. However,
there is a unique extension which is such that variation upon a gauge
transformation only depends on the gauge field, and not on the field
$g$ \cite{Witten:1992mm}.
The resulting gauged WZNW action becomes
\begin{equation}
\begin{split}
  \label{eq:gWZNWaction}
  I(g,A)= I_{WZNW}(g) + \frac{2}{4\pi}\int d^2\!z \Tr\Bigl(&
  A_{L,\bar{z}}\partial g g^{-1} -A_{R,z}g^{-1}\bar{\partial}g
  -A_{L,\bar{z}}gA_{R,z}g^{-1} 
  \\ &
  +\half(A_{L,z}A_{L,\bar{z}}+A_{R,z}A_{R,\bar{z}})
  \Bigr)\ ,
\end{split}
\end{equation}
where $A^i_a$ are the gauge fields, $t^i_L$ are the left
acting generators, $t^i_R$ are the right acting generators, 
and we have used the notation
$A_L = A^i_a t^i_Ld\sigma^a = A_{L,a}d\sigma^a$ and
$A_R = A^i_a t^i_R d\sigma^a = A_{R,a}d\sigma^a$.

The coordinates we shall use for $SL(2,\IR)$ and $SU(2)$ are
\cite{Johnson:2004zq}: 
\begin{equation}
  g_1 =\frac{1}{\sqrt{2}}\begin{pmatrix}
    e^{t_+/2}(x+1)^{1/2}      & e^{t_-/2}(x-1)^{1/2}
    \cr e^{-t_-/2}(x-1)^{1/2} & e^{-t_+/2}(x+1)^{1/2}
  \end{pmatrix} \in SL(2,\IR)\ ,
  \label{eq:parameters1}
\end{equation}
\begin{equation}
  g_2 =\begin{pmatrix}
    \phantom{-}e^{i\phi_+/2}\cos\frac{\theta}{2} 
    & e^{i\phi_-/2}\sin\frac{\theta}{2}
    \cr -e^{-i\phi_-/2}\sin\frac{\theta}{2}
    & e^{-i\phi_+/2}\cos\frac{\theta}{2}\end{pmatrix} \in SU(2)\ ,
  \label{eq:parameters2}
\end{equation}
where $t_\pm=t_L\pm t_R$, and $-\infty\leq t_R, t_L, x\leq\infty$,
and $\phi_\pm=\phi\pm\psi$, $0\leq\theta\leq\pi$,
$0\leq\psi\leq4\pi$, and $0\leq\phi\leq2\pi$. The coordinates
$\theta,\phi,\psi$ are the Euler angles.
Note that $x$ can take any real value here, while remaining in
$SL(2,\IR)$. 
In refs.\cite{Johnson:1994jw,Johnson:1995ga}, 
the range $x=\cosh\sigma\geq 1$ was used. 
The larger range reveals the connection to the Taub and the other NUT
region. This extension is very naturally inherited from the
$SL(2,\IR)$ embedding.

Because the WZ term does not have a gauge-invariant extension, the
above construction does in general have classical anomalies. Only very
certain (essentially left-right symmetric) gaugings produce
gauge-invariant models.
As a way to allow more general gaugings the heterotic coset models
were introduced in ref.\cite{Johnson:1994jw}. In such models,
right-handed fermions are included in order to achieve supersymmetry
in the right sector. These give extra anomalies at one-loop
order. Left-handed fermions are then added, and since we do not
require supersymmetry in the left sector, we can include these with
arbitrary charge. The anomalies from the bosonic sector and the
anomalies from the fermionic sector are of the same form, and
demanding them to cancel, gives algebraic anomaly cancellation
equations which determine the fermion charges in the left sector.
The result is a heterotic string theory free for anomalies.

A problem with the construction at this point is that the anomalies
appear at the classical level in the bosonic sector, while they are a
one-loop effects in the fermionic sector. This makes it hard to
compute the effective spacetime fields in a consistent way.
However, there is an elegant way around this problem, namely by
bosonising the fermions, after which the anomalies are classical
\cite{Johnson:1994jw,Johnson:1995ga}.

Moreover, it has long been known that $D$ fermions upon non-Abelian
bosonisation are given as a WZNW model based on the group $SO(D)$ at
level $1$. So including fermions in the model is done (after
bosonisation) simply by adding an extra WZNW term in the action.

For the model at hand, we need 4 fermions, which will be described by
a bosonic field $g_f\in SO(4)$. Let this group element be
parameterised by
\begin{align} 
  g_f &= \exp\Biggl\{\left( \begin{array}{cc} 
        \Phi_1 \frac{i\sigma_2}{\sqrt{2}} & 
           \\ & -\Phi_2 \frac{i\sigma_2}{\sqrt{2}}
        \end{array} \right)\Biggr\}
    = \left( \begin{array}{cccc}
          \cos\frac{\Phi_1}{\sqrt{2}} & \sin\frac{\Phi_1}{\sqrt{2}} && \\
          -\sin\frac{\Phi_1}{\sqrt{2}} & \cos\frac{\Phi_1}{\sqrt{2}} &&\\
          && \cos\frac{\Phi_2}{\sqrt{2}} & -\sin\frac{\Phi_2}{\sqrt{2}} \\
          && \sin\frac{\Phi_2}{\sqrt{2}} & \cos\frac{\Phi_2}{\sqrt{2}}
        \end{array} \right), 
    \label{eq:parametersf}
\end{align}
where $\Phi_1$ and $\Phi_2$ are $2\pi$ periodic.  

Note that we have effectively gauge-fixed the fermionic sector by only
writing enough fields to fill out an $SO(2)\times SO(2)$ subgroup of
the $SO(4)$. This also anticipates that we will choose the gauge given
in \eqref{eq:ktngaugefix} so as to remove two fields out of the six given
by fully parameterising the $SL(2,\IR)\times SU(2)$, therefore
retaining $\Phi_1$ and $\Phi_2$ in the final model.

So far, everything is identical to what was done in the non-rotating
case \cite{Johnson:2004zq}. The new thing is the implementation of the
gauge symmetry, which is imposed by the equations
\begin{equation}
  \label{eq:ktngauging}
  g_i  \to e^{t^{(i)}_{a,L}\epsilon^a} g_i e^{-t^{(i)}_{a,R}\epsilon^a},
\end{equation}
where $i=1,2,f$, and $a=A,B$.
The gauge generators are now
\begin{equation}
  t^{(1)}_{A,L}=\frac{\sigma_3}{2},
  \quad t^{(1)}_{A,R} =-\delta\frac{\sigma_3}{2},
  \quad t^{(1)}_{B,L}=0,
  \quad t^{(1)}_{B,R}=-\lambda\frac{\sigma_3}{2}\ ,
  \label{eq:generators1}
\end{equation}
\begin{equation}
  t^{(2)}_{A,L}=i\tau\frac{\sigma_3}{2},
  \quad t^{(2)}_{A,R}=0,
  \quad t^{(2)}_{B,L}=0,
  \quad t^{(2)}_{B,R}=-i\frac{\sigma_3}{2}\ ,
  \label{eq:generators2}
\end{equation}
\begin{equation}
\label{eq:generatorsf}
\begin{array}{ll}
  t^{(f)}_{A,L} = \frac{1}{\sqrt{2}}\left(\begin{array}{cccc}
          0 & -Q_A & &\\ Q_A & 0 &&
          \\&& 0 & P_A \\ && -P_A & 0
        \end{array}\right)\ ,
  &
  t^{(f)}_{A,R} = -\frac{1}{\sqrt{2}}\left(\begin{array}{cccc}
          0 & -\delta & &\\ \delta & 0 &&
          \\&& 0 & 0 \\ && 0 & 0
        \end{array}\right)\ ,
  \\ \\
  t^{(f)}_{B,L} = \frac{1}{\sqrt{2}}\left(\begin{array}{cccc}
          0 & -Q_B & &\\ Q_B & 0 &&
          \\&& 0 & P_B \\ && -P_B & 0
        \end{array}\right)\ ,
  &
  t^{(f)}_{B,R} = -\frac{1}{\sqrt{2}}\left(\begin{array}{cccc}
          0 & -\lambda & &\\ \lambda & 0 &&
          \\&& 0 & 1 \\ && -1 & 0
        \end{array}\right)\ .
\end{array}
\end{equation}
For $\tau=0$ we recover the model discussed in
ref.~\cite{Johnson:2004zq}, where the $SU(2)_L$ symmetry is unbroken,
resulting in a spacetime metric with spherical symmetry. For $\tau\neq
0$, the gauging above breaks the $SU(2)_L$ symmetry, and therefore the
metric will not have this symmetry anymore.

As already mentioned, the anomalies can be computed from the gauged
WZNW model. If we do this after the fermions have been bosonised, the
anomaly cancellation equations are given by
\begin{equation}
  \sum_{i,a}
  k_{i}\Tr(t^{(i)}_{a,L}t^{(i)}_{a,L}-t^{(i)}_{a,R}t^{(i)}_{a,R}) 
  = 0, 
\end{equation}
where $i=1,2,f,$ and $a=A,B$.
Written out, this gives
\begin{equation}
\begin{split}
  \label{eq:anomalies}
  k_1(\delta^2-1)-k_2\tau^2 &= 2(Q_A^2+P_A^2-\delta^2),
  \\ k_1\lambda^2 + k_2 &= 2(Q_B^2+P_B^2-(1+\lambda^2)),
  \\ k_1\delta\lambda &= 2(Q_AQ_B+P_AP_B-\lambda\delta),
\end{split}
\end{equation}
where, $k_2=k_1-4$ to give the four-dimensional model a central
charge $c=6$. 
Parameters which satisfy these equations represent meaningful
gauge-invariant theories.
In the following we will write $k_1=k$, $k_2=k-4$.

Note that the $t^{(f)}_R$ are fixed by $(0,1)$ worldsheet supersymmetry,
while in the $t^{(f)}_L$, the $Q_{A,B}$ and $P_{A,B}$ are chosen to cancel
the anomaly {\it via} equation~\eqref{eq:anomalies}.

The gauge fixing is done by imposing\footnote{This follows
  ref.\cite{Johnson:1995ga} rather than ref.\cite{Johnson:2004zq}, but
  the difference is not important as it corresponds simply to a gauge
  transformation on the resulting spacetime.}
\begin{equation}
  t_R=t, \qquad
  t_L=0, \qquad \psi=0. 
  \label{eq:ktngaugefix}
\end{equation}
As was the case in the non-rotating Taub-NUT solution, the gauging
\eqref{eq:ktngauging} also induces a periodicity of the variable
$t_R$, which becomes the time $t$ with the gauge fixing
\eqref{eq:ktngaugefix}.
This can be deduced by studying the action of rotations on the
resulting spacetime metric, but is also clear from the gauging
\eqref{eq:ktngauging}: A $U(1)_B$ transformation with
$\epsilon_B=4\pi$ acts as the identity on the $SU(2)$ space, while in
$SL(2,\IR)$ it translates $t_R\to t_R+4\pi\lambda$. Hence gauging
$U(1)_B$ identifies $t_R\sim t_R+4\pi\lambda$ \cite{Johnson:1995ga}.

\section{Low-energy limit}

This heterotic coset model was constructed in
ref.\cite{Johnson:1995ga}, and the spacetime fields in the large
$k\sim\frac{1}{\alpha'}$ (low-energy) limit were found. The explicit
expressions for the metric and the dilaton are\footnote{We have written
  the extended version where $\cosh\sigma\to x$ and $x$ can take any
  real value. Note also that $\hat{\Phi}$ in this paper corresponds to
  $\frac{1}{4}\hat{\Phi}$ there.}  
\begin{equation}
\begin{split}
  \label{eq:kerrtaubnutlargek}
  ds^2 = & k\Bigl[ \frac{dx^2}{x^2-1}
  -\frac{x^2-1}{(x+\delta-\lambda\tau\cos\theta)^2} 
  (dt-\lambda\cos\theta d\phi)^2 
  \\& \quad +d\theta^2
  +\frac{\sin^2\!\theta}{(x+\delta-\lambda\tau\cos\theta)^2} 
  (\tau dt-(x+\delta)d\phi)^2 
  \Bigr], 
  \\
  e^{2(\hat{\Phi}-\hat{\Phi}_0)} =&(x+\delta-\lambda\tau\cos\theta)^{-\half}.
\end{split}
\end{equation}
This solution has the same Killing horizons at $x=\pm 1$ as the
non-rotating model, and a curvature singularity for
$x=-\delta+\lambda\tau\cos\theta=0$. 
There is also an ergosphere 
outside the horizon 
where the rotational frame dragging makes it impossible for any
particle to remain stationary,
given by $1<x^2<1+\tau^2\sin^2\!\theta$ 
(both in the positive and negative $x$ domains). 
We will come back to these various regions shortly, when discussing the
exact metric. Except for the ergosphere, the overall structure of this
spacetime is similar to the non-rotating stringy Taub-NUT in the
low-energy limit, discussed and illustrated in
ref.\cite{Johnson:2004zq}. 


As in the non-rotating case, there are closed timelike curves in the
NUT regions $x^2>1+\tau^2\sin^2\!\theta$, where  $t$ is timelike and
periodic.  The region $-1<x<1$ is the cosmological Taub patch, where
$t$ is spacelike, and $x$ is timelike. In this region, the
singularities at $x=-1$ and $x=1$ correspond to a Big Bang, and a Big
Crunch respectively. 
 
A crucial observation that was made about the stringy Taub-NUT in the
throat + horizon region, was that it is Misner-like in the
neighbourhood of the horizons $x=\pm 1$.
And for Misner-like spacetimes, there is a semi-classical study
\cite{Hiscock:1982vq} which shows that the vacuum stress-energy tensor
for a conformally coupled scalar field in the background diverges at
the horizons, which indicates an infinite back-reaction that is often
to believed to be such that CTCs are avoided in the exact geometry.
Since the rotation is not essential in this respect, this study is
relevant also in the present case.

\section{Exact metric and dilaton}

The idea \cite{Witten:1991yr} behind computing the spacetime fields is
very simple: 
Integrate out the gauge fields, compare the resulting action to the
nonlinear sigma model, and read off the fields. 
However, if we do this directly, the result is only going to be valid
to first order in the parameter $k$, since the procedure of
integrating out the gauge fields in the naive way is only valid to
first order.
The reason for this is that we are treating  the gauge fields as
classical fields, substituting their on-shell behaviour into the action to
derive the effective nonlinear sigma model action for the rest of the
fields, and ignoring the effects of quantum fluctuations arising at
subleading order in the large $k$ expansion.
To get exact results valid to all orders in $k$, we need to do better.

The exact metric and dilaton
derived from ordinary coset models was first
achieved in ref.\cite{Dijkgraaf:1992ba} in the context
of the $SL(2,\IR)/U(1)$ coset model studied as a model of a
two-dimensional black hole \cite{Witten:1991yr}. In this, and
subsequent studies
\cite{Tseytlin:1991ht,Jack:1993mk,Bars:1992sr,Bars:1993dx}, group
theoretic arguments were applied to write down the exact metric and
dilaton. 

The method we will use in this paper is an alternative one,
which was developed  in refs.\cite{Tseytlin:1993ri,Bars:1993zf}, and
extended to work for heterotic coset models in
ref.\cite{Johnson:2004zq}.
This method exploits the fact that a gauged WZNW model after the
change of variables 
$A_z=\partial_z h_z h_z^{-1}$ and 
$A_{\bar{z}}=\partial_{\bar{z}}h_{\bar{z}}h_{\bar{z}}^{-1}$ 
can be written as a sum of two formally decoupled WZNW models. There
is one for
the field $g'=h_{\bar{z}}^{-1}g h_{\bar{z}}$ at level $k$, and another
for the field $h'=h_z^{-1}h_z$ at level $2c_H-k$, where $c_H$ is
the dual Coxeter number of the subgroup $H$.
Since it is known
\cite{Leutwyler:1992tv,Tseytlin:1993ri,Shifman:1991pa} that the exact
effective action for the WZNW 
model $kI_{WZNW}(g)$, $g\in G$, is simply given by a shift in the level
constant, $S_{eff}=(k-c_G)I_{WZNW}(g)$, where $c_G$ is the dual
Coxeter number of the group $G$, this makes it possible to write down
the exact effective action also for the gauged WZNW model.
Having the effective action, we can change variables back to the
original ones, and integrate out the gauge fields, by solving their
equations of motion and substituting back. This integration is now
exact, since the fields in the effective action are classical.
Finally, we have to take into account that the fermions should be
re-fermionised, so we have to re-write the action into a form which
prepares it for re-fermionisation of the bosonised fermions.

For more detail, we refer the reader to ref.\cite{Johnson:2004zq}
where this has been discussed more thoroughly, including a summary of
how the computations are carried out.
The techniques used for deriving the exact spacetime fields
corresponding to a heterotic coset model can be summarised as follows. 
\begin{enumerate}
  \item Find a Lagrangian formulation of the model in terms of gauged
    WZNW models, where the fermions are included in their bosonised
    form. 
  \item Change to variables where the Lagrangian is a sum of ungauged
  WZNW Lagrangians, and remember to take into account the Jacobian.
  \item Deduce the effective action, which is done merely by shifting
  the level constants.
  \item Change variables back to the original ones. (Note that there
  is no Jacobian this time, as the fields are classical.)
  \item Integrate out the gauge fields.
  \item Prepare the action for re-fermionisation.
  \item Read off the exact spacetime fields.
\end{enumerate}
At the intermediate stages, the calculations produce rather
complicated expressions, and it is a beautiful and highly non-trivial
result that the exact metric takes the simple form
\begin{equation}
\begin{split}
  \label{eq:ktnexactmetric}
  ds^2 =& (k-2)\Bigl[ \frac{dx^2}{x^2-1}
  -\frac{E(x,\theta)}{D(x,\theta)}
  \Bigl(dt+\frac{\Lambda(x,\theta)}{E(x,\theta)}d\phi\Bigr)^2 
  +d\theta^2 +\frac{x^2-1}{E(x,\theta)}\sin^2\!\theta d\phi^2
  \Bigr],
  \\ & E(x,\theta)= x^2-1-\tau^2\sin^2\!\theta,
  \\ & D(x,\theta)=(x+\delta-\lambda\tau\cos\theta)^2
  -\frac{4}{k+2}(x^2-1-\tau^2\sin^2\!\theta),
  \\ & \Lambda(x,\theta)=-\lambda\cos\theta(x^2-1)
  +\tau\sin^2\!\theta(x+\delta). 
\end{split}
\end{equation}
As expected, for $\tau=0$ this reduces to the non-rotating solution of
ref.\cite{Johnson:2004zq}, and for $k\to\infty$ we recover
the low-energy metric \eqref{eq:kerrtaubnutlargek}.

The general expression for the dilaton is
\begin{equation}
  e^{2\Phi} = 
  (\Delta)^{-\frac{1}{2}}(\det\mathcal{G}_{mn})^{-\frac{1}{2}},
\end{equation}
where one part is from integrating out the gauge fields, and the other part
is from re-fermionisation. These are given by
\begin{equation}
\begin{split}
  \Delta(x,\theta) & =\Bigl[ (k-2)px+pq-(2P_A+(k-2)\tau\cos\theta)r\Bigr]^2
  +4(1-\tau^2)\Bigl[ r^2 - p^2 \Bigr],
  \\
  \det \mathcal{G}_{mn} & = 4(k+2)(k-2)^3\frac{D(x,\theta)}{\Delta(x,\theta)},
\end{split}
\end{equation}
where $p$, $q$ and $r$ are defined as
\begin{equation}
  \label{eq:defpqr}
  p = k-2+2P_B, \quad
  q=(k+2)\delta +2Q_B, \quad
  r=(k+2)\lambda +2Q_B.
\end{equation}
This gives the exact dilaton
\begin{equation}
  \label{eq:ktnexactdilaton}
  \hat{\Phi}-\hat{\Phi}_0=  -\frac{1}{4}\ln D(x,\theta),
\end{equation}
which is again a pleasingly simple expression to result from
remarkable cancellations of complicated factors.

\subsection{Properties of metric}

The metric \eqref{eq:ktnexactmetric} has two Killing vectors,
$\xi=\frac{\partial}{\partial t}$ and
$\psi=\frac{\partial}{\partial\phi}$ representing time translation
symmetry, and axial symmetry. 
A Killing horizon is defined as the surface where a linear
combination of the Killing vectors becomes null
\cite{Wald:1984rg}. 
The singularities at $x=\pm 1$ are such horizons, since the
Killing vector
\begin{equation}
  \chi = \frac{\partial}{\partial t} +
  \Omega_H\frac{\partial}{\partial\phi};
  \qquad \Omega_H=\frac{\tau}{\delta\pm 1},
\end{equation}
becomes null at $x=\pm 1$:
\begin{equation}
  \chi^2 =G_{tt} + 2G_{t\phi}\Omega_H +
  G_{\phi\phi}\Omega_H^2 
  = \frac{k-2}{D}(\tau-(\delta\pm 1)\Omega_H)^2
  =0,
\end{equation}
for the above chosen value of the parameter $\Omega_H$,
which is interpreted as the angular velocity at the
horizon and is proportional to $\tau$ as anticipated.
Note that the horizon is independent of the value of all the
parameters in the model, so it is the same as in the low-energy limit,
and also the same as in the non-rotating case.
Note also that the angular velocity of the horizon, $\Omega_H$ is
independent of $k$.

\begin{figure}[ht]
  \begin{center}
	\mbox{
		\includegraphics[width=6cm]{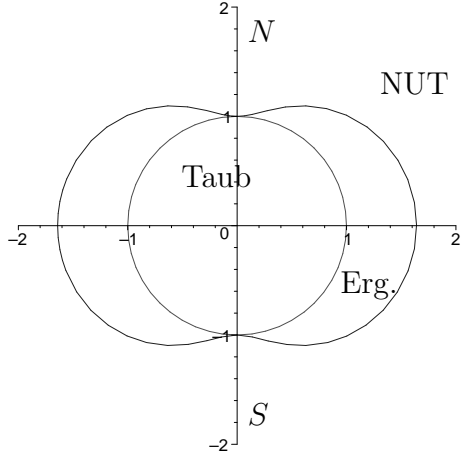}
		\put(-80,155){$N$}
		\put(-80,10){$S$}
		\put(-105,100){Taub}
		\put(-30,135){NUT}
		\put(-45,60){Erg.}
	}
  \end{center}
  \caption[Horizon and ergosphere of stringy Kerr-Taub-NUT geometry.
    ]{\small Polar plot of the locus of the horizon (inner circle) and
    ergosurface (outer deformed circle) of the stringy Kerr-Taub-NUT
    spacetime. The radial direction is $\sqrt{|x|}$, and the angle
    $\theta$ runs from $0$ at the north pole (N) to $\pi$ at the south
    pole (S). (This plot is for $\tau=2.5$.)}
  \label{fig:horizon}
\end{figure}

The metric component $G_{tt}$ becomes zero when $E(x,\theta)=0$ \ie, when
\begin{equation}
  x^2=1+\tau^2\sin^2\!\theta ~~\in~ [1,1+\tau^2].
\end{equation}
This surface, which we shall call the ergosurface, lies outside the
horizon (see figure \ref{fig:horizon}). Between the horizon and this
ergosurface is the ergosphere, a region where no stationary particles
can exist. 
To see that the ergosphere really is a region where particles cannot
be stationary, consider the following. Assume that a particle follows
a trajectory with tangent vector $u$, which has to be timelike \ie,
$u^2<0$. 
In the stationary case, the only motion is in the time direction, so
the tangent vector is $u=u^t \frac{\partial}{\partial t}$, 
where $u^t=\frac{dt}{ds}$, and $s$ is proper time along the curve. 
But this gives $u^2=G_{tt}(u^t)^2>0$ in the ergosphere (where
$G_{tt}>0$), and so the assumptions are inconsistent: No stationary
motion is possible in the ergosphere. 
What happens instead is that particles are affected by the
\emph{rotational frame dragging} and inevitably follow the rotation of
the black hole (as observed from infinity) \cite{Wald:1984rg}.
%
%
%

Note that both the Killing horizons and the ergosurface are
independent of $k$, and are therefore not not modified by the
high-energy corrections of the spacetime.

\subsubsection{Curvature singularities}

The metric \eqref{eq:ktnexactmetric} is ill defined for
$D(x,\theta)=0$, which is also  a true curvature singularity.  
This can be verified by computing the curvature. It is
then seen that both the Ricci scalar and the Kretschmann scalar
($R^{\mu\nu\rho\sigma}R_{\mu\nu\rho\sigma}$) behave 
like $\sim \frac{1}{D(x,\theta)^2}$, and so indeed $D=0$ represents a
curvature singularity. (The same argument would also show that $x^2-1=0$ or
$E=0$ are \emph{not} curvature singularities.)

First of all, notice that $D=0$ only has solutions if 
\begin{equation}
  x^2-1-\tau^2\sin^2\!\theta=E(x,\theta)>0,
\end{equation}
that is, singularities are only found outside the
ergosphere. (Inside the ergosurface, $D$ is always positive.)
The equation $D(x,\theta)=0$ solved for $x$ gives
\begin{equation}
\begin{split}
  x = x_\pm(\theta) = &-\frac{1}{k-2}(\beta\mp\sqrt{\alpha^2}),
  \\& \alpha^2 = 4(k+2)(\delta-\lambda\tau\cos\theta)^2
  -4(k-2)(1+\tau^2\sin^2\!\theta) ,
  \\& \beta = (k+2)(\delta-\lambda\tau\cos\theta).
	\label{eq:ktnsingularities}
\end{split}
\end{equation}
\begin{figure}[ht]
  \begin{center}
	\mbox{
		\includegraphics[width=6cm]{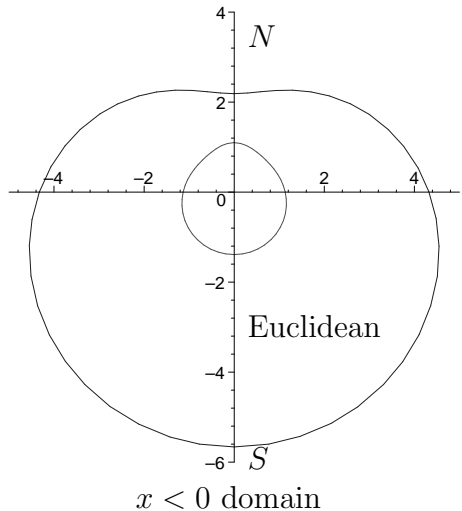}
		\put(-80,160){$N$}
		\put(-80,0){$S$}
		\put(-80,50){Euclidean}
	}
	\\
	$x<0$ domain
  \end{center}
  \caption{\small Polar plots of the locus of the singularities for the
    small $\tau$ case. The radial direction is $\sqrt{|x|}$, and the
    angle $\theta$ is zero at the north pole (N) and $\pi$ at the south
    pole (S). There is symmetry in the $\phi$ direction.
    (This plot is for $k=3,\delta=2,\lambda=14,\tau=0.1$.)}
  \label{fig:singsmalltau}
\end{figure}
\begin{figure}[ht]
  \begin{center}
  \begin{tabular}{ccc}
  \mbox{
	\includegraphics[width=6cm]{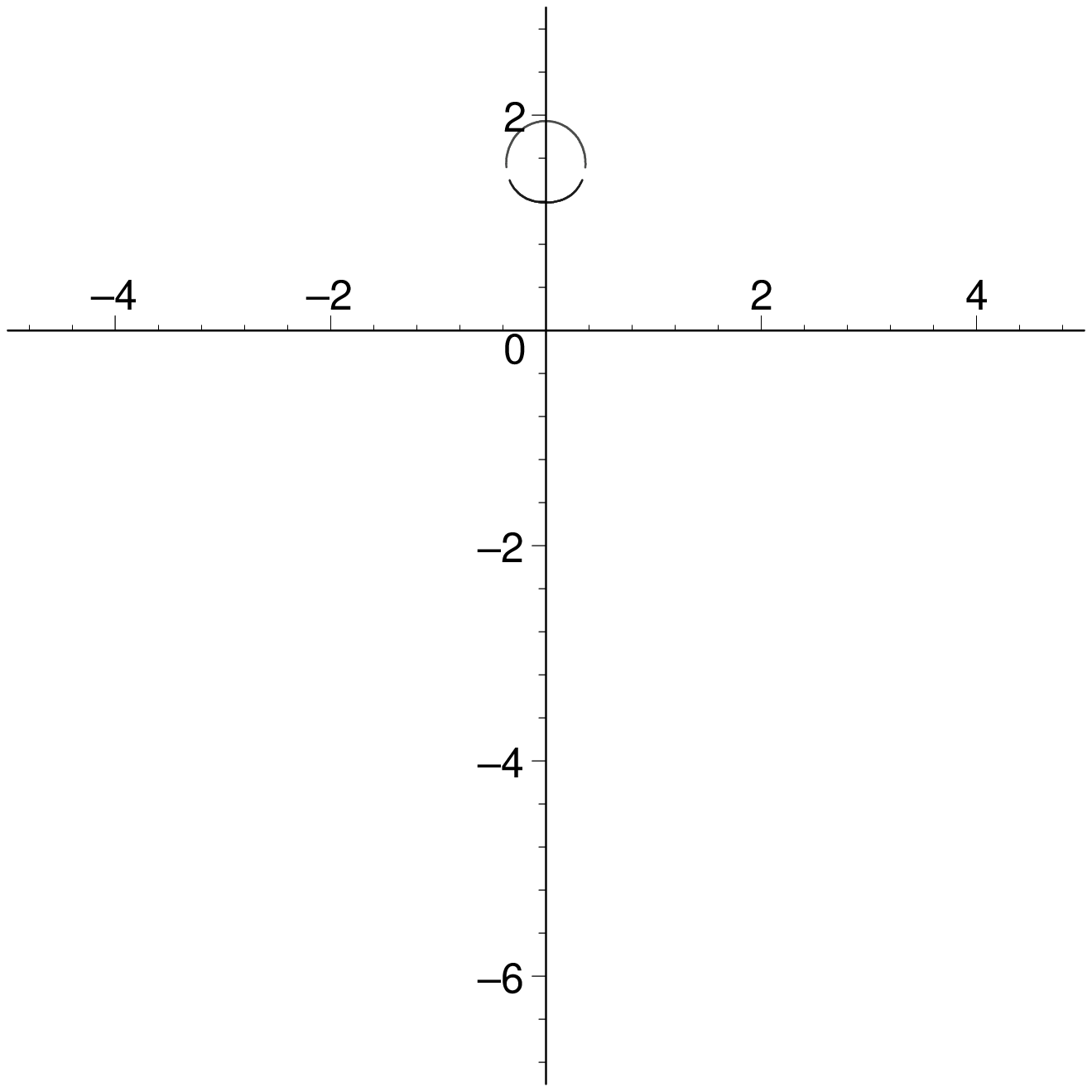}
	\put(-80,160){$N$}
	\put(-80,5){$S$}
	\put(-55,150){Euclidean}
	\put(-60,153){\vector(-4,-1){25}}
	}
  &&
  	\mbox{
		\includegraphics[width=6cm]{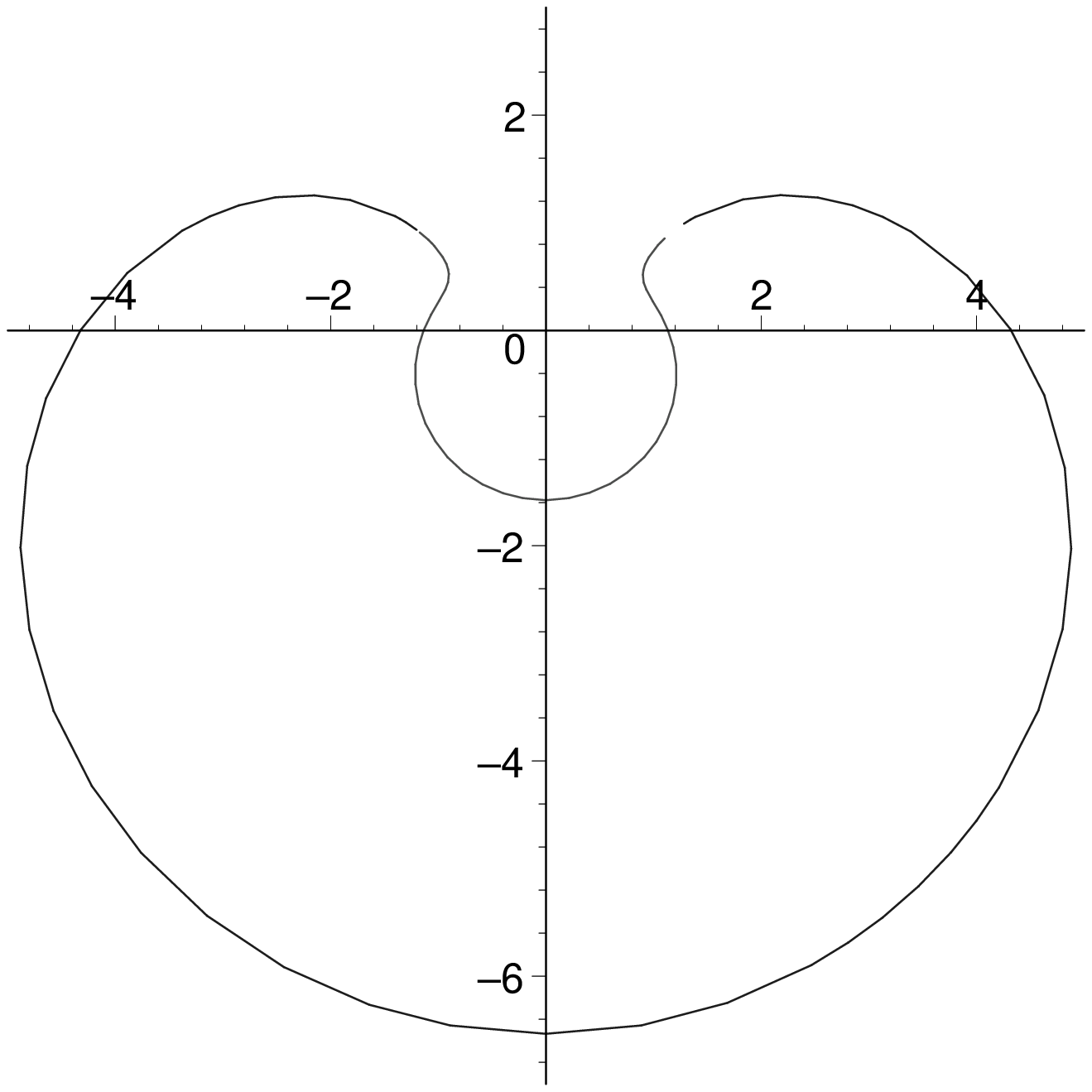}
		\put(-80,160){$N$}
		\put(-80,0){$S$}
		\put(-120,60){Euclidean}
	}
  \\
  $x>0$ domain && $x<0$ domain
  \end{tabular}
  \end{center}
  \caption[Singularities in stringy Kerr-Taub-NUT geometry, large
    $\tau$]{\small Polar plots of the locus of the singularities for the
    large $\tau$ case. 
    (This plot is for $k=3,\delta=2,\lambda=14,\tau=0.18$.)}
  \label{fig:singlargetau}
\end{figure}
For given values of the parameters and for $\theta$, the solutions $x_\pm$
have the same sign (which is the same sign as $-\beta$), as can easily
be seen from the following:
\begin{equation}
  \beta^2-\alpha^2
  = (k+2)(k-2)(\delta-\lambda\tau\cos\theta)^2 +
  4(k-2)(1+\sin^2\!\theta)
  \ge 0.
\end{equation}
Hence, $|\beta|\ge|\alpha|$.

The singularities $x_\pm(\theta)$ can be divided into two classes,
depending on the values of the parameters $k,\lambda,\delta,\tau$. 
%
%
Assuming $k,\delta,\lambda$ are given, there is a critical value for
$\tau$ above which $\alpha^2$ in equation \eqref{eq:ktnsingularities}
is \emph{not} positive definite. Call this critical value
$\tau_{c}(k,\delta,\lambda)$.  Then ``small rotation'' means values
$\tau<\tau_{c}$, and ``large rotation'' means values $\tau>\tau_{c}$. 

The first class is the \emph{small rotation case}, where
$\tau<\tau_c$. In this
case $\alpha^2$ is always positive, and $x_\pm$ exist for all
$\theta$, and are both negative. This situation is illustrated in
figure \ref{fig:singsmalltau}. The spacetime structure in this case is
a smooth deformation of the non-rotating stringy Taub-NUT spacetime
discussed in ref.\cite{Johnson:2004zq}, with the same essential
features. 
The singularities appear in the negative $x$ region, and enclose a
region of Euclidean signature. Seen from the positive $x>1$ NUT
region, and the from the Taub region $-1<x<1$ this is a singularity
hidden behind a horizon, while from the negative $x<-1$ NUT regions
they appear as naked singularities. 

The second class is the \emph{large rotation case}, where
$\tau>\tau_c$. In this case, $\alpha^2$ becomes negative for
some values of $\theta$. Hence, for these angles, there are no
divergences. This is illustrated in figure \ref{fig:singlargetau}. 
What happens is that the two surfaces $x\pm(\theta)$ connect and form
a ``bubble'' outside the ergosphere. One such bubble is centred at
the south pole ($\theta=\pi$) and appear in the negative $x$
region. This also makes the two NUT regions in the negative $x$ domain
merge together into one connected region. 
Another bubble may or may not appear at the north pole ($\theta=0$)
for positive $x$. 
This is rather different from the non-rotating case, and quite exotic
behaviour. The bubbles still enclose regions of Euclidean signature,
but since they appear in both the positive and the negative $x$
region, all the NUT regions are plagued by naked singularities.

\section{Discussion}

The model discussed in this paper is a generalisation of the stringy
Taub-NUT spacetime of ref.\cite{Johnson:2004zq}. The rotational symmetry is
broken in the general case. 
It was demonstrated in that earlier work that the $\alpha'$ corrections
do not modify the spacetime significantly with regards to the CTCs in
the non-rotating case, and this result persists in the rotating case. So
all the comments made there carry on to the more general model of this
paper. This is a valuable observation
in that it shows us that the results of ref.\cite{Johnson:2004zq}
are not simply a coincidence happening only for that particular
spacetime. Noting the miraculous cancellation that 
gave the simple form for the exact metric, we could have been tempted
to believe there was something very special happening in that case.
Now, as we see the same happening again, an interpretation of it as a
mere coincidence seems even more unlikely. 
A more reasonable interpretation of the mild $\alpha'$ corrections
near the horizons seems to be that string theory really does not rule
out the possibility of CTCs. This view has already been discussed in
ref.\cite{Johnson:2004zq}. 

The above comments are closely related to the observations that the
horizons and the ergosurface are not modified by the high-energy
corrections of the spacetime. 

The curvature singularities in the present model differ from what we
saw in the non-rotating case, and this deserves a comment. First of
all, it is important to keep in mind that the dilaton blows up at the
singularity, so the string coupling $g_s$ is in no sense small. Hence,
$g_s$ corrections may completely alter the geometry at the
``would-be-singularities'' where $D(x,\theta)=0$. How to compute these
corrections, however, is beyond reach with our technology at present.
So the exotic singular structure of the metric
\eqref{eq:ktnexactmetric} might only be an artifact of working in the
classical limit (which is a good approximation only if $g_s\to 0$). 

If we ignore this for a moment, we have spacetimes containing
problematic naked singularities in the NUT regions. If the rotation is
small, these appear only for negative $x$, and so we could still make
sense of the positive $x$ NUT region since it would be protected from
the singularities by the horizons at $x=\pm 1$. In this case 
the Taub region has a natural extension past $x=1$ into the region
$x>1$, giving a cosmology with a post Big Crunch scenario. 
If the rotation is large, on the other hand,
the singularities appear both in the positive
and negative $x$ NUT regions, and any sensible extension of the Taub
region seems impossible.

%

\section*{Acknowledgements}

I am grateful to Clifford Johnson for suggesting this study, and for
useful conversations. I also thank Bill Spence, Douglas Smith, and
James Gray for useful comments.
This work is funded by the Research Council of Norway through a
doctoral student fellowship.

\bibliographystyle{JHEP-2}
\bibliography{../durhame}

\providecommand{\href}[2]{#2}\begingroup\raggedright\begin{thebibliography}{10}

\bibitem{Johnson:2004zq}
C.~V. Johnson and H.~G. Svendsen, {\it An exact string theory model of closed
  time-like curves and cosmological singularities},
  \href{http://arXiv.org/abs/hep-th/0405141}{{\tt hep-th/0405141}}. Submitted
  to \textit{Phys. Rev. D.}

\bibitem{Taub:1951ez}
A.~H. Taub, {\it Empty space-times admitting a three parameter group of
  motions},  {\em Annals Math.} {\bf 53} (1951) 472--490.

\bibitem{Newman:1963yy}
E.~Newman, L.~Tamubrino and T.~Unti, {\it Empty space generalization of the
  schwarzschild metric},  {\em J. Math. Phys.} {\bf 4} (1963) 915.

\bibitem{Johnson:1995ga}
C.~V. Johnson and R.~C. Myers, {\it A conformal field theory of a rotating
  dyon},  {\em Phys. Rev.} {\bf D52} (1995) 2294--2312
  [\href{http://arXiv.org/abs/hep-th/9503027}{{\tt hep-th/9503027}}].

\bibitem{Johnson:1995kv}
C.~V. Johnson, {\it Heterotic coset models},  {\em Mod. Phys. Lett.} {\bf A10}
  (1995) 549--560 [\href{http://arXiv.org/abs/hep-th/9409062}{{\tt
  hep-th/9409062}}].

\bibitem{Johnson:1994jw}
C.~V. Johnson, {\it Exact models of extremal dyonic {4-D} black hole solutions
  of heterotic string theory},  {\em Phys. Rev.} {\bf D50} (1994) 4032--4050
  [\href{http://arXiv.org/abs/hep-th/9403192}{{\tt hep-th/9403192}}].

\bibitem{Galtsov:1994pd}
D.~V. Galtsov and O.~V. Kechkin, {\it Ehlers-harrison type transformations in
  dilaton - axion gravity},  {\em Phys. Rev.} {\bf D50} (1994) 7394--7399
  [\href{http://arXiv.org/abs/hep-th/9407155}{{\tt hep-th/9407155}}].

\bibitem{Johnson:1994ek}
C.~V. Johnson and R.~C. Myers, {\it {Taub--NUT} dyons in heterotic string
  theory},  {\em Phys. Rev.} {\bf D50} (1994) 6512--6518
  [\href{http://arXiv.org/abs/hep-th/9406069}{{\tt hep-th/9406069}}].

\bibitem{Kallosh:1994ba}
R.~Kallosh, D.~Kastor, T.~Ortin and T.~Torma, {\it Supersymmetry and stationary
  solutions in dilaton axion gravity},  {\em Phys. Rev.} {\bf D50} (1994)
  6374--6384 [\href{http://arXiv.org/abs/hep-th/9406059}{{\tt
  hep-th/9406059}}].

\bibitem{Witten:1992mm}
E.~Witten, {\it On holomorphic factorization of {WZW} and coset models},  {\em
  Commun. Math. Phys.} {\bf 144} (1992) 189--212.

\bibitem{Hiscock:1982vq}
W.~A. Hiscock and D.~A. Konkowski, {\it Quantum vacuum energy in taub - nut
  (newman-unti-tamburino) type cosmologies},  {\em Phys. Rev.} {\bf D26} (1982)
  1225--1230.

\bibitem{Witten:1991yr}
E.~Witten, {\it On string theory and black holes},  {\em Phys. Rev.} {\bf D44}
  (1991) 314--324.

\bibitem{Dijkgraaf:1992ba}
R.~Dijkgraaf, H.~Verlinde and E.~Verlinde, {\it String propagation in a black
  hole geometry},  {\em Nucl. Phys.} {\bf B371} (1992) 269--314.

\bibitem{Tseytlin:1991ht}
A.~A. Tseytlin, {\it On the form of the black hole solution in d = 2 theory},
  {\em Phys. Lett.} {\bf B268} (1991) 175--178.

\bibitem{Jack:1993mk}
I.~Jack, D.~R.~T. Jones and J.~Panvel, {\it Exact bosonic and supersymmetric
  string black hole solutions},  {\em Nucl. Phys.} {\bf B393} (1993) 95--110
  [\href{http://arXiv.org/abs/hep-th/9201039}{{\tt hep-th/9201039}}].

\bibitem{Bars:1992sr}
I.~Bars and K.~Sfetsos, {\it Conformally exact metric and dilaton in string
  theory on curved space-time},  {\em Phys. Rev.} {\bf D46} (1992) 4510--4519
  [\href{http://arXiv.org/abs/hep-th/9206006}{{\tt hep-th/9206006}}].

\bibitem{Bars:1993dx}
I.~Bars and K.~Sfetsos, {\it $sl(2,\mathrm{R})\times su(2) / \mathrm{R}^2$
  string model in curved space-time and exact conformal results},  {\em Phys.
  Lett.} {\bf B301} (1993) 183--190
  [\href{http://arXiv.org/abs/hep-th/9208001}{{\tt hep-th/9208001}}].

\bibitem{Tseytlin:1993ri}
A.~A. Tseytlin, {\it Effective action of gauged wzw model and exact string
  solutions},  {\em Nucl. Phys.} {\bf B399} (1993) 601--622
  [\href{http://arXiv.org/abs/hep-th/9301015}{{\tt hep-th/9301015}}].

\bibitem{Bars:1993zf}
I.~Bars and K.~Sfetsos, {\it Exact effective action and space-time geometry in
  gauged {WZW} models},  {\em Phys. Rev.} {\bf D48} (1993) 844--852
  [\href{http://arXiv.org/abs/hep-th/9301047}{{\tt hep-th/9301047}}].

\bibitem{Leutwyler:1992tv}
H.~Leutwyler and M.~A. Shifman, {\it Perturbation theory in the
  wess-zumino-novikov-witten model},  {\em Int. J. Mod. Phys.} {\bf A7} (1992)
  795--842.

\bibitem{Shifman:1991pa}
M.~A. Shifman, {\it Four-dimension aspect of the perturbative renormalization
  in three-dimensional chern-simons theory},  {\em Nucl. Phys.} {\bf B352}
  (1991) 87--112.

\bibitem{Wald:1984rg}
R.~M. Wald, {\em General Relativity}.
\newblock Chicago, USA: Univ. Pr., 1984.
\newblock 491~p.

\end{thebibliography}\endgroup

\end{document}